\documentclass[12pt]{iopart}

\usepackage{epsfig}
\usepackage{graphicx}
\begin{document}

\title[Superconducting THz metasurfaces for ultrastrong light-matter coupling ]{Superconducting complementary metasurfaces for THz ultrastrong light-matter coupling}

\author{Giacomo Scalari, Curdin Maissen}
\address{Institute of Quantum Electronics, ETH Z\"urich, Switzerland}
\author{Sara Cibella, Roberto Leoni}
\address{CNR-IFN Rome, Italy}
\author{Pasquale Carelli}
\address{DIEI, Universit\`a dell'Aquila, L'Aquila,  Italy}
\author{ Federico Valmorra, Mattias Beck, J\'er\^ome Faist }
\address{Institute of Quantum Electronics, ETH Z\"urich, Switzerland}

\ead{scalari@phys.ethz.ch}

\begin{abstract}
A superconducting metasurface operating in the THz range and based on the complementary metamaterial approach is discussed. Experimental measurements as a function of temperature and magnetic field display a  modulation  of the metasurface with a change in transmission amplitude and frequency of the resonant features.   Such a metasurface is successively used as a resonator for a cavity quantum electrodynamic experiment displaying ultrastrong coupling to the cyclotron transition of a 2DEG. A finite element modeling is developed and its results are in good agreement with the experimental data. In this system a normalized coupling ratio of $\frac{\Omega}{\omega_c}=0.27$ is measured and a clear modulation of the polaritonic states as a function of the temperature is observed. 
\end{abstract}

\maketitle

\section{Introduction: ultrastrong light-matter coupling with THz metasurfaces and 2-dimensional electron gas}
Light-matter interaction phenomena lie at the heart of quantum photonics \cite{Haroche:book:06}. Recently, there has been a considerable progress in the understanding and in the realization of light-matter coupling experiments covering a large portion of the electromagnetic spectrum, from the visible to the microwave range.  When considering low frequency photons, such experiments can benefit from the excellent characteristics of superconducting cavities which ensure low loss rates allowing observation and manipulation of polaritonic systems, from atomic physics to solid state.  \cite{Haroche:RMP:01, Walraff:Nat:04}

\begin{figure}[h]
\begin{center}
   \includegraphics[width=120mm]{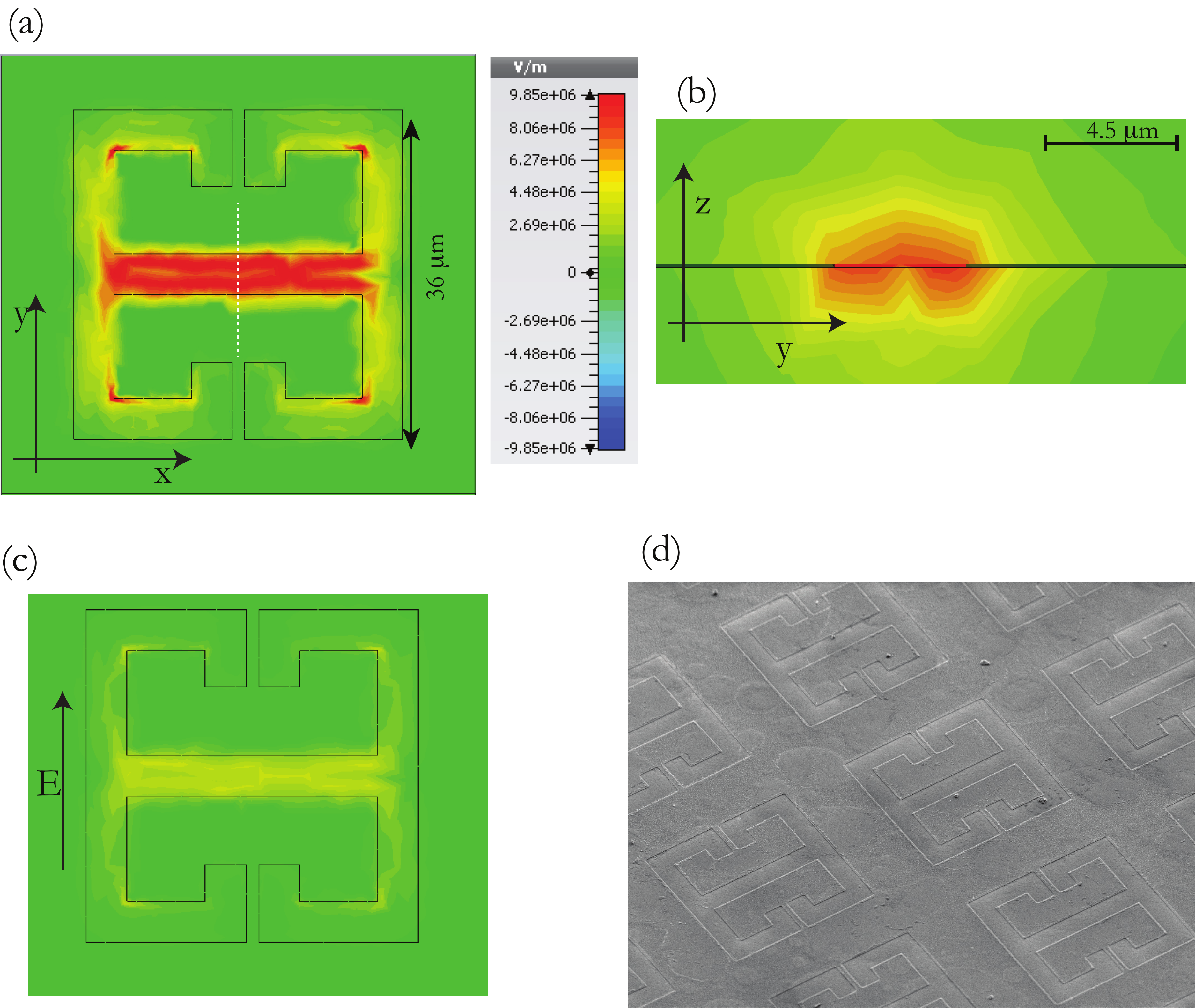}
    \caption{(a): In-plane electric field intensity $E_{plane}=\sqrt{E_x^2+E_y^2}$ for the resonator operating at 440 GHz. The simulation is performed for the Nb in the supercoducting state, with the measured values for the superconductor complex surface impedance, reported in Fig. \ref{resexperiment}(b). (b): in-plane electric field distribution in the y-z plane following the white dashed line of panel (a). (c): in-plane electric field intensity for the resonator operating in the normal state. (d): SEM picture of the fabricated sample with 100 nm thick Nb metasurface on top of the 2DEG. } 
\label{resscheme}
   \end{center}
\end{figure}

Light-matter coupling physics in solid state system has been recently approaching a new regime, called ultrastrong coupling \cite{Ciuti:PRB:05:115303-1,Niemczyk-NATPHYS-2010} , where the Rabi frequency of the system $\Omega$ is comparable to the bare excitation frequency of the matter part $\omega$. Semiconductor-based systems operating in the Mid-IR \cite{anapparaPRB,Guenter:Nature:2009,Delteil:PRL:12} and THz \cite{Geiser:PRL:12,todorov:PRL:2010,muravev:prb:2013} are particularly attractive for the study of this peculiar regime because very large dipole moments can be achieved and the system can benefit from the enhancement of the light-matter coupling by $\sqrt{N}$ deriving form the simultaneous coupling of N electrons with the same photonic mode of the cavity.
We recently demonstrated that the cyclotron transition of a 2-dimensional electron gas (2DEG) coupled to a metasurface of  subwavelength split-ring resonators can attain the ultrastrong coupling regime showing record high values of the normalized light-matter coupling ratio $\frac{\Omega}{\omega_c}=0.58$ \cite{scalari:science:2012,scalari:jap:2013}. This system is very promising for probing the ultimate limits of the ultrastrong light-matter coupling since the coupling ratio scales with the filling factor $\nu$ of the 2DEG. This means that, as long as the cyclotron transition can be resolved, the experiments can be scaled down in frequency achieving coupling ratios in principle much higher than unity \cite{Hagenmuller:2010p1619}.

The ultrastrong coupling regime has been predicted to display intriguing and peculiar quantum electrodynamics features:  Casimir-like \cite{Wilson:Nat:11} squeezed vacuum photons upon either non-adiabatic change or periodic modulation in the coupling energy \cite{Ciuti:PRB:05:115303-1},  non-classical radiation from chaotic sources \cite{Ridolfo:PRL:13}, ultrafast switchable coupling \cite{Ridolfo:PRL:11}, spontaneous conversion from virtual to real photons \cite{Stassi:PRL:13}. 
The paper is organized as follows: in Sec.\ref{metamaterial} we present the metasuface design and its fabrication. In Secs.\ref{metamaterialTemp} and \ref{metamaterialBfield}  we analyze the behavior of the metasurface as a function of the temperature and of the applied magnetic field respectively, together with finite element modeling. In Sec.\ref{metamaterialstrongC} we then present the measurements were the Nb metasurface is strongly coupled to the cyclotron transition of a 2DEG.

\section{The metamaterial cavity: fabrication, experimental results and modeling }\label{metamaterial}

In our previous experiments  \cite{scalari:science:2012,scalari:jap:2013}, we employed metasurfaces of split-ring resonators that, when probed in transmission, show an absorption dip at the resonant frequency of the LC mode. The characteristic anticrossing behavior of the polaritonic system is probed by scanning the magnetic field. This provides a linear tuning of the cyclotron energy. With such experimental arrangement we have observed three absorption features \cite{scalari:science:2012,scalari:jap:2013}: two are related to the light-matter coupled system and the third middle peak is the cyclotron  signal coming from  the material which lies in between the resonators. This cyclotron signal is only weakly coupled to the metasurface and follows the expected linear dispersion for a cyclotron transition $\hbar \omega_c=\frac{\hbar e B}{m^*}$. In order to observe with greater detail the spectroscopic features of the polaritonic branches, we  now change the configuration of the metamaterial cavity and we employ a complementary cavity. 

As already shown in several papers \cite{Falcone:PRL:04,chen:OE:07} the complementary metamaterial is obtained by  exchanging the roles of the vacuum areas and of the metals composing the metasurface. The resulting metasurface is constituted of a metallic sheet with openings  with the shapes of the resonators. When excited with an electric field complementary to the one used in the case of the direct metamaterial (i.e. along the y axis, perpendicular to the central gap as in Fig.\ref{resscheme}(c)), the resonator will display a transmission spectrum which is complementary to the one shown by the split ring resonators. 

\begin{figure}[h]
\begin{center}
   \includegraphics[width=100mm]{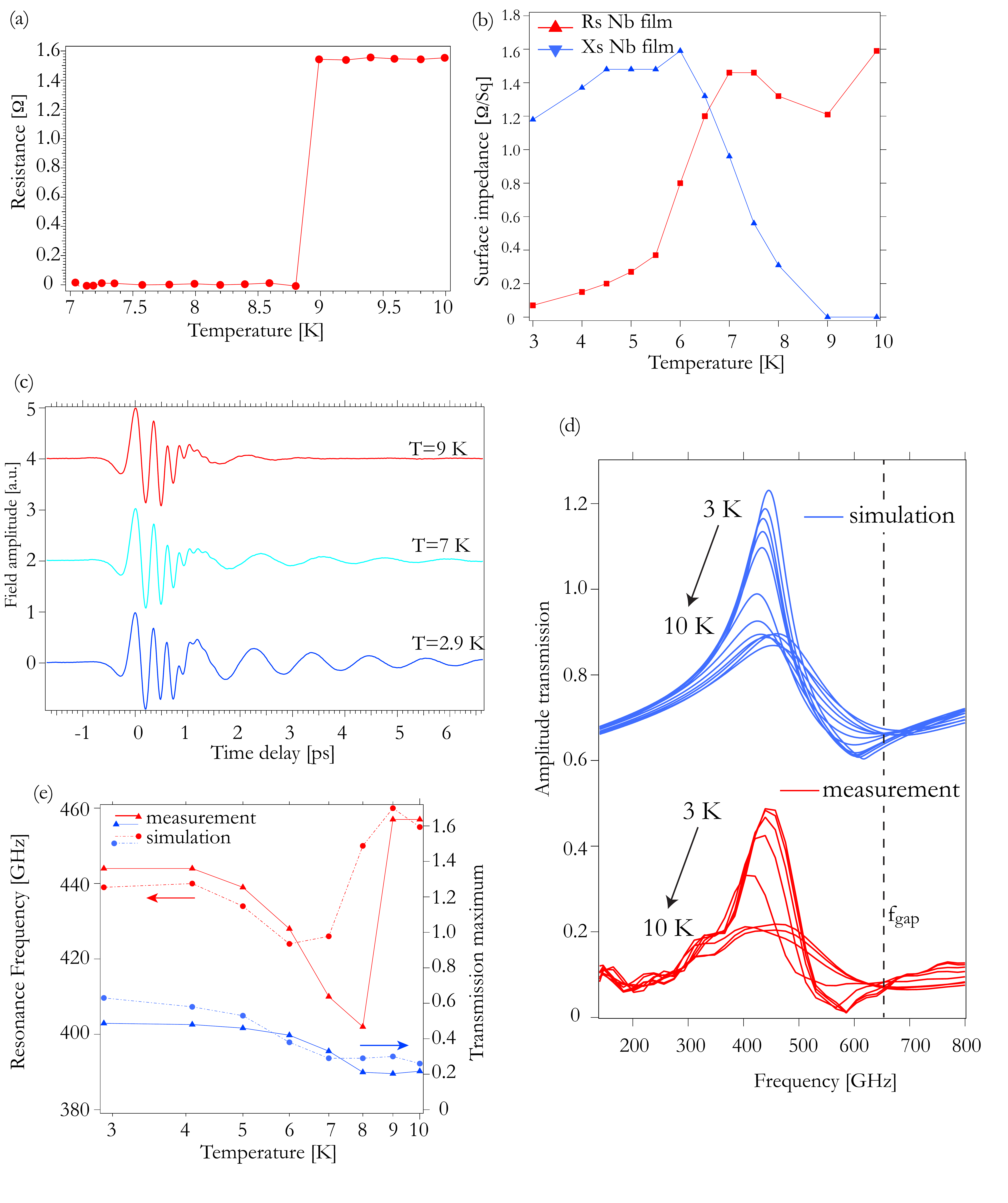}
    \caption{(a): DC resistance measurement as a function of the temperature for the 100 nm thick Nb film used to fabricate the metasurface. (b): measured complex surface impedance at the frequency of 440 GHz for the Nb film as a function of temperature. (c): time traces for the metasurface deposited on SI GaAs for three different  temperatures.  (d): simulated (top) and measured (bottom)  transmission for the superconducting metasurface deposited on the SI GaAs as a function of the temperature varied in the range 2.9-10 K. The surface impedance values Z$_s$=R$_s$+i X$_s$ are extracted from the conductivity measurement of the Nb film reported in panel (b) of this figure. (e): measured (triangles) and simulated (circles) transmission peak and frequency as a function of the temperature for the Nb metasurface. } 
\label{resexperiment}
   \end{center}
\end{figure}

The intriguing quantum optical predictions for an ultrastrongly coupled system rely on a non-adiabatic modulation \cite{Ciuti:PRB:05:115303-1,Guenter:Nature:2009,Wilson:Nat:11} of  the system's parameters; in our case on timescale faster than the Rabi frequency of the system (100-400 GHz).  The fabrication of a superconducting cavity offers an interesting opportunity in this direction, since the cavity characteristics strongly depend on the state (superconducting or normal) in which the material operates. On the other hand, even if in our case the quality factor of the split-ring resonator is radiatively limited, the presence of the superconductor will mitigate the ohmic losses yielding a longer polariton coherence time. In the last few years many examples of superconducting GHz and THz metamaterials have been presented, both using BCS superconductors like Nb and NbN \cite{Ricci:APL:05,Jin:OE:10,Zhang:OE:12}, or dealing with hight T$_c$ superconductors \cite{Chen:PRL:10}. For our cavity we chose Nb, which is a well-known superconducting material widely employed in THz science. The niobium film of the resonators, about 100 nm thick, is deposited by dc-magnetron sputtering both on a semi insulating (SI) GaAs (for the control sample) and on a sample containing a  triangular well 2DEG (for the strong-coupling experiment). On top of the Nb film we spun an electronic resist (Polymethyl Methacrylate, positive tone electronic resist) and the pattern was defined by the direct writing with the electron beam lithography. Successively, the niobium is selectively removed by means of a dry reactive ion etching. A  scanning electron microscope picture of the fabricated devices is presented in Fig.\ref{resscheme}(d). The Nb film has a critical temperature Tc of 8.7 K, as results from DC resistance measurements reported in Fig.\ref{resexperiment}(a). From this T$_c$ value we can infer a gap value of $E_g=2\Delta=4.1 K_B T_c\simeq 3$ meV which corresponds to $f_{gap}\simeq 730$ GHz \cite{Pronin:PRB:98}. 

 In  Fig. \ref{resscheme}(a) we report the design of our resonator (very similar to the one reported in Ref. \cite{chen:OE:07} ) and the in-plane electric field distribution (which is the one relevant for the inter-Landau level coupling) for the LC resonance when the superconductor is in the superconducting state. More details about the modeling of the metasurface wil be given in Sec.\ref{metamaterialTemp}. The capacitor, where the majority of the electric field is concentrated, is now constituted by the central section of the resonator of width 4.5 $\mu m$. The effective volume of the cavity in this case can  be estimated as $V_{cav}=27.5 \times 4.5 \times 3.2 \times 10^{-18}=8.6\times 10^{-16}$ m$^3$ which yields a ratio of $\frac{V_{cav}}{(\lambda/2n_{eff})^3}\simeq 1 \times 10^{-3}$.

In  Fig. \ref{resscheme}(c) we report the in-plane electric field distribution simulated with the superconductor in the normal state: the electric field enhancement results approximately one half than the one simulated in the superconducting case.
% two metamaterial configurations: if we compare the effective interaction area per resonator we find an increase by almost a factor of 4 for the complementary metamaterial. 
 %In the same figure we compare the transmission signal for both metasurfaces for a sample including one modulation doped triangular quantum well which constitutes the %matter part of our light-matter experiment. 
\subsection{THz metasurface characteristics as a function of the temperature}\label{metamaterialTemp}

The complementary metasurface, first fabricated on SI GaAs,  is then probed with THz-Time Domain Spectroscopy (TDS) in a range of temperatures above and below T$_c$. The experimental setup employed in this paper is based on a THz-TDS system coupled to a split-coul superconducting magnet and is described in detail in Refs. \cite{scalari:science:2012,scalari:jap:2013} As visible from Fig. \ref{resexperiment}(c,d,e), a clear change in the metasurface Q factor and a shift of the resonance frequency is observed between 8 K an 9 K, as already observed in direct metamaterials fabricated with other kind of superconductors \cite{Chen:PRL:10}.  The deduced high-frequency T$_c$ of 8.5 K is in good agreement with the DC value and the one found in similar structures \cite{Jin:OE:10}. 
%The same metasurface has been investigated at a temperature of 3.5 K as a function of the applied magnetic field. As visible from the data reported in Fig. TTT, the quality factor and the resonant frequency present respectively a degradation and a redshift until a field value of B=1.45 T where a smooth transition occurs and the metasurface tunes back to the value of 450 GHz (CHECK ) with a Q factor of ZZZ. The behavior is similar to what observed when changing the temperature above and below the critical T$_c$. 
%  The smooth transition is expected for a type II superconductor as Nb \cite{Tinkham:96}. 

The observed behavior of the superconducting complementary metasurface can be explained by considering the complex conductivity of the metasurface and the LC nature of the resonance. 
For a superconductor like Nb in the normal state (above T$_c$) the conductivity $\sigma$ is essentially real and, in thin films has been measured in the range of $\sigma_{T>T_c}=2.5 \times 10^7$ S/m \cite{Pronin:PRB:98}. Below $T_c$, $\sigma$ is mainly complex and frequency dependent: the complex part displays low values in proximity of the superconducting gap frequency $f_{gap}$.
In order to model the system we can proceed  following Chen et al,\cite{Chen:PRL:10} by  considering the surface impedance of a superconducting film with a thickness d can be expressed as $Z_s=R_s+iX_s=\sqrt{\frac{i\omega \mu_0}{\sigma}}Coth(d\sqrt{{i\omega \mu_0 \sigma}})$ where $\sigma$ is the complex conductivity. The resistive part $R_s$ and the reactive part $X_s$ can then be connected to the  parameters of the resonator seen as an RLC circuit. 

 Still along the analysis presented in  Ref.\cite{Chen:PRL:10} we can  express the resonance of the LC mode in the following way:

\begin{equation}
\nu_{res}=\frac{1}{2\pi}\sqrt{\left( \frac{1}{(L_g+L_k)C}-\frac{R^2}{4(L_g+L_k)^2}   \right)}
\end{equation}

where $L_g$ is the geometrical value of the inductance, $L_k$ represents the kinetic inductance due to the imaginary part of the conductivity and R is the resistance of the circuit. When the temperature is above $T_c$, the value of $L_k$ is extremely small and the resistance R is inversely proportional to the real part of the Nb conductivity.
The shift of the resonance for temperatures just below $T_c$ is mainly due to the emergence of the imaginary part of the conductivity which effectively introduces the kinetic inductance of the Cooper pairs. Across the superconducting transition , we will have the value of R that will change due to the change of the real part of the conductivity and the reactive part which will become relevant introducing a supplementary term L$_k(X_s)$ which effectively lowers the frequency. This explains the redshift of the resonance between 9 K and 7 K.  Further reduction of the temperature increases the conductivity, thus reducing the resistance. This is reflected in two aspects, a narrowing of the resonance due to reduced dissipation and a blue shift due to the reduction of R which restores an higher value for the resonance frequency . 
% vanishing value of the real part of the conductivity, which affects the resistive part of the complex impedance. Then, following the two-fluid model, the density of Cooper pairs increases as the temperature is further lowered below T$_c$ and the kinetic inductance of the pairs account for the rising value of the reactive part in the complex impedance.
 
We measured the complex conductivity of the employed 100 nm thick Nb unstructured film as a function of the temperature with THz-TDS. Successively  we calculated the complex surface impedance and we extracted the values at the resonant frequency f=440 GHz. These values are reported in Fig.\ref{resexperiment}(b) and  then used to perform the 3D calculation. 
  As visible from the Fig.\ref{resexperiment}(d,e)  the experimental results are well reproduced  by the 3D modeling performed with CST microwave studio using a surface impedance model for the superconductor.    
The small  discrepancy between the simulated amplitude transmission and the measured one   has already been observed in similar experiments  \cite{Chen:PRL:10} and can be ascribed to different causes.  We believe the main reason is that the values for the complex conductivity are considered as frequency-independent in the simulations: this is not true and they present quite large variations especially when approaching the gap frequency: our resonator is in fact operating slightly below the gap ($\nu_{res}=440 GHz < f_{gap}=730$ GHz) and this simplification can explain the semi-quantitative agreement between simulations and experiments.  It is remarkable that no relevant features are observed in the metamaterial spectra in correspondence and above of the gap frequency f$_{gap}$.

\subsection{THz metasurface characteristics as a function of perpendicular magnetic field}\label{metamaterialBfield}

\begin{figure}[h]
\begin{center}
   \includegraphics[width=130mm]{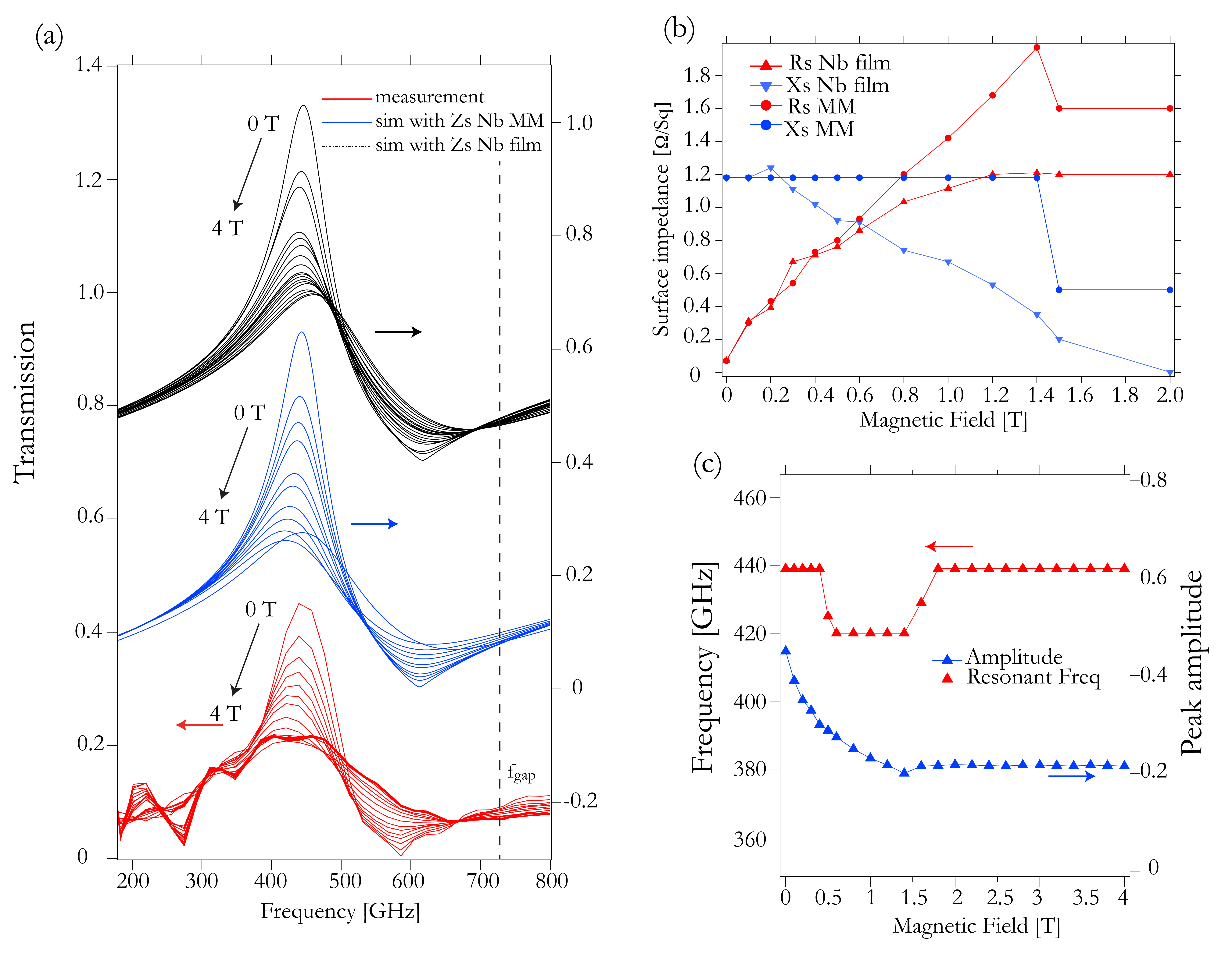}
    \caption{(a):  simulated (top, dashed black and continuous ) and measured (bottom) transmission for the superconducting metasurface fabricated on the SI GaAs as a function of the magnetic field at a temperature of 2.9 K. The surface impedance values Z$_s$ Nb film are extracted from the conductivity measurement of the Nb film reported in panel (b) of this figure. (b): measured complex surface impedance at the frequency of 440 GHz for the Nb film as a function of the applied magnetic field (points labeled Nb film). (c): Measured peak amplitude and resonant frequency for the Nb metasurface as a function of magnetic field at T=2.9 K} 
\label{resBfield}
   \end{center}
\end{figure}

In order to model the complete system when the 2DEG is coupled to the resonator and measured as a function of the magnetic field we need then study and model the resonator behavior as a function of the applied magnetic field. 
We now analyze the data of the superconducting metasurface at a temperature of 2.9 K subjected to a perpendicular magnetic field of increasing strength. The measured transmission spectra are  reported in Fig.\ref{resBfield}(a)(red traces). 
The application of a perpendicular magnetic field leads to a steady decrease of the transmission at resonance and a slight redshift until a field value of 1.5 T. At this point a transition occurs and the resonator linewidth broadens considerably together with a blueshift of the resonant frequency: such a shape and both the transmission value and the resonant frequency stay  constant until high values of the applied magnetic field (see Fig. \ref{resBfield} (c)).

 It is known that in a type II superconductor such as Nb, the application of a perpendicular magnetic field produces current 
vortices in the sample  which surround regions of normal state conductivity, where the magnetic field can penetrate the sample \cite{Tinkham:96,Bardeen:PR:65, Ricci:07}.
%yielding for the total inductance in presence of a magnetic field an  expression of the kind $L_{Bfield}=L_g+L_k+L_v$. 
At the same time, there is an increasing portion of the sample which presents normal state for Nb, and these  two effect enter in a non trivial way into the expression for the surface impedance. The plain Nb film was measured this time as a function of the magnetic field at a constant temperature of T=2.9 K. The extracted surface impedance at 440 GHz is reported in Fig.\ref{resBfield} (b) and is used to model the resonator. The results are reported in Fig.\ref{resBfield}(a) as black lines: the change in transmittance is well reproduced but the frequency shift shows an opposite behavior. In order to have a numerical model to use in the simulation of the complete system, we then deduce effective values for the metasurface complex impedance parameter which show good agreement with the experimental data. These parameters are reported in Fig.\ref{resBfield} (b) and labeled as R$_s$ MM and X$_s$ MM since are effective values for the metamaterial. The results of the simulations with such parameters are reported in Fig.\ref{resBfield} (a)(blue lines). In order to reproduce the experimental results, we need to keep the reactance value X$_s$ constant until the critical field value of 1.5 T and change only the resistive part. This behavior, which differs from what experimentally observed with the measurements of the Nb film, can be qualitatively explained by the presence of a structured surface on the micrometric scale (the metamaterial itself) that alters the behavior and the vortex formation,\cite{Baert:PRL:95}  leading, in the case of the metasurface, to a different evolution of the surface impedance as a function of the applied magnetic field.

%The progressive redshift of the resonance as a function of the applied magnetic field can then be explained by the increasing weight of the term $L_v$. When the $H_{c2}$ is reached, the metamaterial is not superconducting anymore and we indeed observe a change in the transmission amplitude and in the resonance frequency, at about 1.4 T. The critical field for bulk Nb is 1.9 T. The resonance in fact presents a much broader peak above 1.4 T and other spectral features of the transmission present an abrupt change at the same magnetic field value (see for example the low frequency feature at 140 GHz).

\begin{figure}[h]
\begin{center}
   \includegraphics[width=130mm]{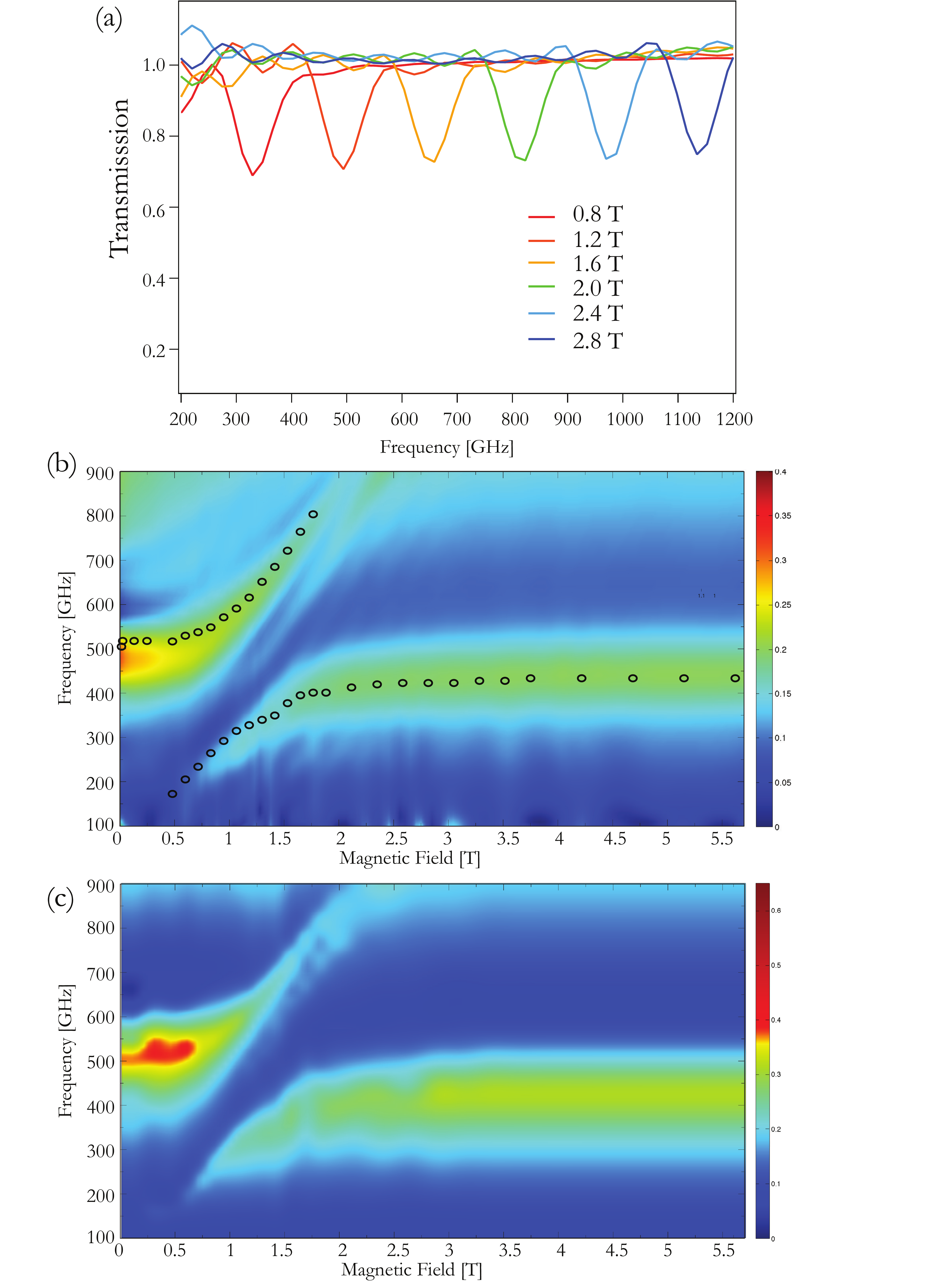}
    \caption{(a): Cyclotron absorption from the triangular well 2DEG for different values of the applied magnetic fields at a temperature of 2.9 K.(b): Color plot of sample transmission a a function of the applied magnetic field at a temperature T=2.9 K. The black circles are the transmission maxima extracted from the numerical simulation of panel (c).  (c): Simulated transmission as a function of the applied magnetic field. } 
\label{strongcouplingcolorplot}
   \end{center}
\end{figure}

\section{Ultrastrong coupling with superconducting metasurface}\label{metamaterialstrongC}

Now we consider the complete system and we analyze the measurements where the superconducting metasurface is strongly coupled to the Landau levels of the 2DEG: we use a  triangular quantum well of sheet carrier density $\rho_{2DEG}=3.2 \times 10^{11}$ cm$^{-2}$ (measured with Hall bars) with the channel lying 100 nm below the surface.
An example of the inter-Landau level transition, or cyclotron resonance, that we observe from the triangular 2DEG without any metasurface is reported in Fig.\ref{strongcouplingcolorplot}(a).

 The resonator we use displays field enhancement also in the TM polarization (electric field perpendicular to the semiconductor surface), which has been used to observe light-matter coupling with intersubband  transitions \cite{dietze:OE:11}. From measurements (not shown) of the 2DEG alone in a tilted magnetic field \cite{Schlesinger:PRL:83} we  deduce an energy separation between the first two subbbands of our triangular quantum well of at least 3 THz,  so the ISB does not couple to the TM field of our resonator for the LC mode. 

In Fig.\ref{strongcouplingcolorplot}(b) is visible a color plot of a transmission experiment carried out at 2.9 K. Clear anticrossing between upper and lower branch is observed and we can measure a normalized coupling ratio $\frac{\Omega}{\omega}=0.27$ which makes the system operating in the ultrastrong light-matter coupling regime. Due to the use of a complementary metasurface the polaritonic branches are especially clear because the cyclotron signal is not present in between the two branches. A transmission maximum parallel to the cyclotron dispersion is observable starting from 1.4 T and a frequency of 0.5 THz and 
is due to the coupling of the second mode of the resonator to the cyclotron resonance. We can model then the complete system using, as cavity parameters, the ones obtained with the measurements described in the previous paragraph and labeled as R$_s$ MM and X$_s$MM.
 As visible from the color plot of Fig. \ref{strongcouplingcolorplot}(c), the experimental data is well reproduced by the numerical modeling. For convenience, the transmission peaks deduced from the simulations have been reported on the experimental data of Fig.\ref{strongcouplingcolorplot}(b) as black circles. The model is able to reproduce also weak features as the coupling with the second mode of the cavity which is clearly visible as a weak maximum of transmission both in the direct measurements and in the simulation.  
The 2DEG can be experimentally characterized by a complex conductivity, measured again by THz-TDS as in Ref. \cite{Wang:OL:07}. From the conductivity $\sigma_{2DEG}$ we can deduce  the frequency dependent complex dielectric constant $\epsilon=\epsilon_{\infty}+\frac{i \sigma_{2DEG}}{\epsilon_0 \omega L_{eff}}$.  Such expression can be used to perform finite element simulations and contains all the physics of the system in a semiclassical description.

%and yields the correct description also for $B\rightarrow 0$, where the dielectric constant  value is lowered by the free carrier contribution, which generates the blue shift of the cavity mode observed at B=0.
%On the other limit, for high values of the magnetic field , the cavity resonant frequency assumes the value measured on the control sample, were no free electrons are present.
The 2DEG is modeled by having an effective thickness of 200 nm (the value resulting from band structure calculations is 20 nm, which is too small to yield consistent results with the 3D FE solver).
As already observed in our previous experiments \cite{scalari:science:2012,scalari:jap:2013}, the bare resonator frequency $\omega_0$ is blue-shifted to $\omega_0+\Delta \omega$ when the cavity is loaded with the 2DEG electrons which are free to move in the x-y plane without applied magnetic field. The magnitude of the shift is directly related to the normalized coupling strength and is due to the quadratic terms of the light-matter Hamiltonian \cite{Ciuti:PRB:05:115303-1,Hagenmuller:2010p1619}. As discussed also in Refs. \cite{todorov:PRL:2010,Todorov:PRB:12,Delteil:PRL:12,Geiser:PRL:12} the presence of this polaritonic gap is a signature of the ultrastrong light-matter coupling regime. By taking the value at B=0 for the upper polariton branch and the value for $B \rightarrow \infty$
for the lower polariton branch as expressed from the solution of the secular equation of Ref.\cite{Hagenmuller:2010p1619} we obtain the following equation that relates the normalized light-matter coupling ratio $\frac{\Omega}{\omega_c}$ to the normalized frequency gap $\frac{\Delta \omega}{\omega_0}$:

\begin{eqnarray}
\frac{\Omega}{\omega_c}=\frac{1}{2}\sqrt{\left(\frac{\Delta \omega}{\omega_0}+1\right)^2-1}
\end{eqnarray}

If we take the experimental data of the shift of the resonance frequency at B=0 T and we compute the expected light-matter coupling ratio we find 0.28 which is in excellent agreement with the value  0.27 directly deduced from the anticrossing splitting $2\Omega$.
We notice that the shift $\Delta \omega$ is also quantitatively reproduced by the finite element model (the value of the resonance at B=0 in Fig.\ref{strongcouplingcolorplot}(b) is higher than the one at B=5.7 T ). This happens because the conductivity model used for the 2DEG yields the correct description also for $B\rightarrow 0$, where the dielectric constant value is lowered by the free carrier contribution, which generates the blue shift of the cavity mode observed at B=0.
On the other limit, for high values of the magnetic field, the cavity resonant frequency assumes the value measured on the control sample (SI GaAs substrate), were no free electrons are present.

The value of the light-matter coupling ratio is independent from the material used to fabricate the cavities as it depends only on the geometry of the metamaterial elements. We performed similar strong coupling measurements on standard Au cavities  of identical design on the same heterostructure and  we obtain $\frac{\Omega}{\omega_c}$=0.27, in excellent agreement with what found with Nb cavities. A comparison with the respective direct metamaterial will be carried out elsewhere \cite{maissen:unpub:2013}.  
The cavity design of the present work is the complementary version of what published in Ref. \cite{scalari:jap:2013}: in that case the normalized coupling ratio was slightly higher ($\frac{\Omega}{\omega_c}=0.34$) because the volume of the cavity  for the direct metasurface is smaller,  as well as the mode extension in the growth direction which affects the coupling strength.

\begin{figure}[h]
\begin{center}
   \includegraphics[width=110mm]{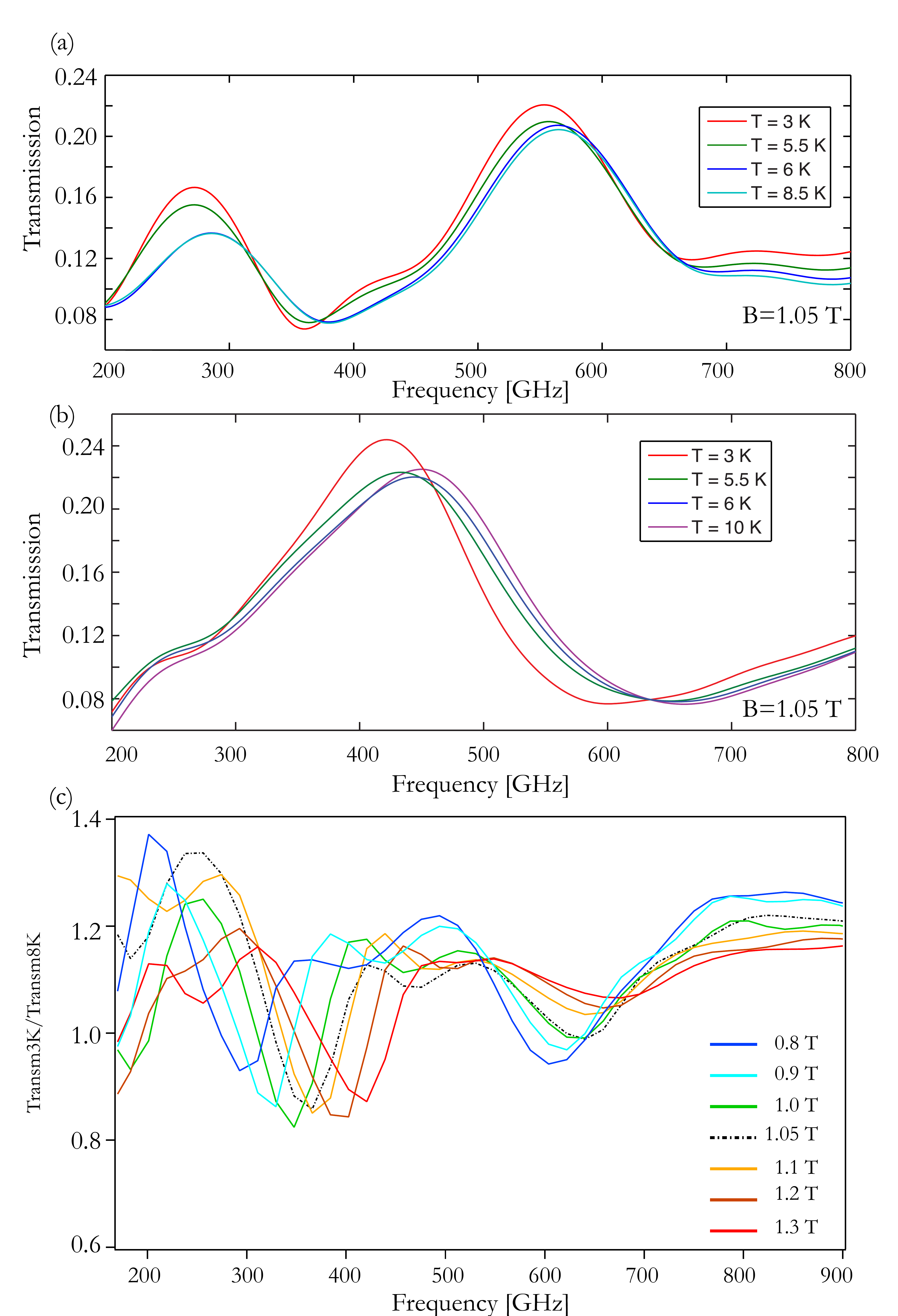}
    \caption{(a): Sample transmission at the anticrossing field of 1.05 T as a function of temperature. (b): Transmission for the super conducing metasurface deposited on SI GaAs. (c): ratio between transmittance spectra at 3 K and 8 K for different magnetic fields ranging from 0.8 T to 1.3 T. } 
\label{strongcouplingtemp}
   \end{center}
\end{figure}

%old caption peak transmission as a function of the temperature for upper and lower polaritons at B=0.8 T before the resonance, at B=1.05 T at resonance and at B=1.3 T after resonance together with the peak transmission for the reference resonator on SI GaAs at the resonant field. (c): frequencies for the two polariton branches and the reference SI GaAs NB resonator as a function of the temperature.

The possibility to modulate the ultrastrong coupling regime by changing the characteristics of the cavity is illustrated in Fig.\ref{strongcouplingtemp}(a) where we show a series of transmission spectra taken at the resonant field B=1.05 T by changing the sample's temperature from below to above T$_c$. We can clearly observe a modulation both in intensity and in frequency of both polariton peaks, which reflects what observed in the same conditions for the cavity only, reported in Fig.\ref{strongcouplingtemp}(b). 
In order to better analyze the impact of the superconducting transition on the ultrastrongly coupled system, we plot in Fig.\ref{strongcouplingtemp}(c) the ratio between the transmission spectra at 3.5K to the ones at 8 K (above the superconducting transition) for magnetic fields values in the range 0.8-1.3 T. As visible from the graph, the relative change of both polaritonic branches is significant, reaching 35 \% for the lower branch and 20 \% for the upper branch. It is interesting to note the different behavior of the lower and upper branches as a function of the applied field. The relative change for the upper branch steadily decreases to zero as long as the magnetic field is increased. On the contrary, the relative change of the lower branch reaches a maximum for the resonant value of B=1.05 T but never goes below 15 \% in the range of magnetic fields 0.8-1.3 T. This can be explained by the fact that the upper branch becomes more and more matter-like and then the impact of the temperature change is much weaker compared to the cavity. The lower branch, on the contrary, is becoming more cavity-like and thus is more affected by the cavity change.
%If we analyze the temperature dependence of the complete system including metasurface and 2DEG and we compare it with the behavior of the resonator only, we can gain insight into the strong coupling mechanism. We report in Fig.\ref{strongcouplingtemp}(b,c) the amplitude and frequency for the spectral features of the complete sample and of the reference resonator on SI GaAs as a  function of the temperature and the applied magnetic field. The clear switching displayed by the cavity as a function of the temperature is progressively smoothened by the coupling with the 2DEG which bears a much weaker temperature dependence in the range 3.5-10 K. If we follow the frequency change for the strong coupling sample from below the resonant field (0.8 T) to the resonant field at 1.05 T and above resonance (1.3 T) we can see a "redistribution" of the weight of the resonator features which shift from the upper branch to the lower one which becomes more cavity like as the magnetic field is increased. 

\section{Conclusions} 

In conclusion we presented light-matter coupling experiments in the THz range employing complementary superconducting metasurfaces. The system operates in the ultrastrong coupling regime and the presence of the superconducting cavity allows the modulation of the resonant frequency and of the quality factor of the resonator. As already shown in literature, the superconducting metasurface  can be modulated  in a  an ultrafast way, on a sub picosecond timescale \cite{Singh:Nanophot:12}. In future work we will  leverage on that aspect and realize non-adiabatic experiments in the ultra strong coupling regime in our magnetopolaritonic system. 
%
%\textit{Methods: The finite element 3 D simulation of the system were performed with CST microwave studio. The 2DEG was modeled as a an electric gyrotropic medium assuming a plasma frequency $\omega_{p}=\sqrt{\frac{\rho_{2DEG}e^2}{L_{qw}m^*\epsilon_0}}=2.76\times 10^{13}$ rad/s and a collision frequency of $\omega_{coll}=2\times 10^{11}$ rad/s. The thickness of the layer used in the simulations was $L_{qw}=$200 nm, as deduced from, band structure simulations, and the channel of the 2DEG starts 150 nm below the semiconductor surface. The superconducting metasurface was simulated with a surface impedance model by employing the measured  values for the reactance and resistance as a function of the temperature. The temperature dependence of the complex conductivity above and below T$_c$ has been analyzed in detail experimentally in Ref. \cite{Pronin:PRB:98}. }

\ack
This research was supported by the Swiss National Science Foundation (SNF) through the National Centre of Competence in Research Quantum Science and Technology and through the SNF grant n. 129823.
G.S. would like to acknowledge F. Chiarello for discussions.

\section*{References}
%\bibliography{references}

\end{document}